# Wide-field high-resolution microscopy via high-speed galvo scanning and real-time mosaicking

Ziyi Huang[*a], Rosalinda Xiong[a], Yaning Wang[a], Jinglun Yu[a], Jin U. Kang[a]
[a]Dept. of Electrical and Computer Engineering, Johns Hopkins Univ., 3400 N. Charles Street, Baltimore, MD USA 21218

## ABSTRACT

Wide-field high-resolution microscopy requires fast scanning and accurate image mosaicking to cover large fields of view without compromising image quality. However, conventional galvanometric scanning, particularly under sinusoidal driving, can introduce nonuniform spatial sampling, leading to geometric inconsistencies and brightness variations across the scanned field. To address these challenges, we present an image mosaicking framework for wide-field microscopic imaging that is applicable to both linear and sinusoidal galvanometric scanning strategies. The proposed approach combines a translation-based geometric mosaicking model with region-of-interest (ROIs) based brightness correction and seam-aware feathering to improve radiometric consistency across large fields of view. The method relies on calibrated scan parameters and synchronized scan–camera control, without requiring image content–based registration. Using the proposed framework, wide-field mosaicked images were successfully reconstructed under both linear and sinusoidal scanning strategies, achieving a field of view of up to 2.5 × 2.5 cm² with a total acquisition time of approximately 6 s per dataset. Quantitative evaluation shows that both scanning strategies demonstrate improved image quality, including enhanced brightness uniformity, increased contrast-to-noise ratio (CNR), and reduced seam-related artifacts after image processing, while preserving a lateral resolution of 7.81 μm. Overall, the presented framework provides a practical and efficient solution for scan-based wide-field microscopic mosaicking.

## 1. INTRODUCTION

Wide-field high-resolution microscopy plays an important role in histological analysis, materials inspection, and biomedical research [1]. However, conventional microscopic systems are inherently limited by the field of view of the objective lens, making it difficult to directly acquire continuous centimeter-scale images while maintaining high spatial resolution. As a result, achieving large-area microscopic imaging with high resolution, geometric consistency, and brightness uniformity remains a fundamental challenge. A commonly adopted strategy is to combine the fast scanning with image mosaicking, in which multiple individual fields of view (FOVs) are acquired across the sample and registered within a unified coordinate system to extend the effective imaging area. However, under microscopic imaging conditions, weak texture, limited signal-to-noise ratio, and sensitivity to illumination variations make the mosaicking outcome highly sensitive to scanning trajectory stability and the robustness of the image processing pipeline against geometric and radiometric inconsistencies [2].

In high-speed galvo-based microscopic scanning systems, galvanometric mirrors are typically driven using either sinusoidal or linear waveforms. Sinusoidal scanning is widely employed in laser scanning, confocal microscopy, and optical coherence tomography [3-4] due to its mechanical smoothness, but its inherently nonuniform angular velocity leads to nonuniform spatial sampling along the scan direction, resulting in positional deviations and brightness variations [5]. In contrast, linear scanning can theoretically provide more uniform spatial sampling and more predictable geometric behavior [6], though it suffers from lower scanning efficiency caused by the necessary turnaround time at each line end For both scanning methods, practical implementations are constrained by the bandwidth and acceleration limits of galvo hardware, and require calibration or computational remapping to achieve accurate spatial positioning.

In wide-field microscopic mosaicking, these scan-induced geometric and radiometric nonidealities can accumulate across multiple frames, significantly degrading seam continuity and global brightness uniformity [7]. In particular, when reliable image features for registration are unavailable, mosaicking accuracy relies heavily on scan-parameter-based geometric modeling, followed by brightness correction and seam suppression. Therefore, a systematic evaluation of scanning strategies and the integration of robust image processing methods are essential for high-quality wide-field mosaicking under high-speed scanning conditions.

[*]zhuang85@jh.edu; phone 1 410 963-4983

## 2. METHODS

### 2.1 System setup and scanning geometry

The experimental system (Fig. 1) utilizes a 2D galvo scanner (GVS002, Thorlabs) integrated with a microscope objective (LSM-04, Thorlabs) and zoom lens (zoom 7000, Navitar) to perform wide-field scanning. A 1951 USAF resolution target (R1L1SIP, Thorlabs) was used as the specimen at a 42 mm working distance. Each raw frame (1000 x 1000 pixels) covers 5 x 5 mm FOV under constant magnification and focus.

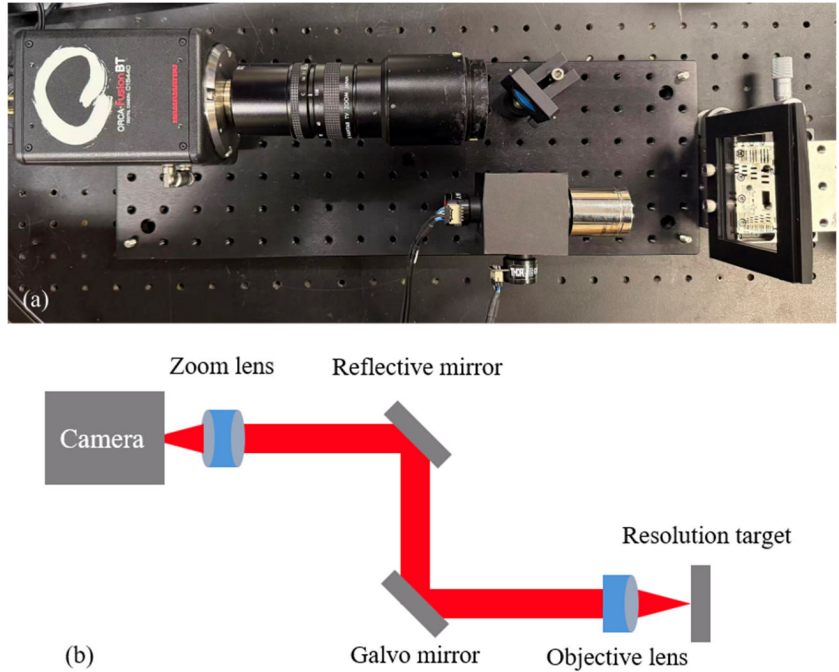

Figure 1. Experimental imaging system and corresponding optical layout. (a) Photograph of the experimental imaging system. (b) Schematic illustration of the optical configuration.

During image acquisition, synchronization between galvo scanning and camera imaging was achieved via software-based timing control. The system paused for a 30-ms settling period at each coordinate to allow NI-DAQ driven galvo mirrors to stabilize before retrieving r the most recent frame from the camera buffer and mapping into the current scan coordinate. Therefore, each acquired frame was uniquely associated with a predefined scan coordinate.

Given the fixed magnification and working distance, we assume individual frames were free of field-dependent distortions, allowing the mosaicking model to formulate primarily as a translation-based linear model.

To compare scanning strategies, we acquired images using . $N_{row} = 10$ rows and $N_{col} = 10$ columns. In both scanning modes, the Y direction was driven by uniformly spaced voltage steps ($\Delta V_Y$=1.1 V). Under linear scanning, the X-axis galvo was driven with a constant voltage step ($\Delta V_X$=1.1 V), resulting in uniform spacing between adjacent sampling positions. Based on system calibration, the pixel-to-voltage conversion factor was determined as $S_X = 402$ pixels/V and $S_Y = 468$ pixels/V for the X and Y axes. During mosaicking, the position of each image frame was determined by its row and column indices $(i, j)$ in the scanning sequence and mapped into a unified global coordinate system. The translation in the global mosaic coordinate system was calculated as

$$\Delta x = j \cdot \Delta V_X \cdot S_X \quad (1)$$
$$\Delta y = i \cdot \Delta V_Y \cdot S_Y, \quad (2)$$

Under sinusoidal scanning, the X-axis scanning positions were parameterized by a sinusoidal function of the scan index,

$$V_X(j) = V_0 + A \cdot \sin\left(\frac{(j-1)\pi}{N_{col}-1} - \frac{\pi}{2}\right), \quad (3)$$

leading to position-dependent variations in the spatial displacement between adjacent frames. In this case, the initial position of each image frame in the global coordinate system was determined from the corresponding scan voltage $V_X(j)$ through the pixel-to-voltage mapping,

$$\Delta x = V_X(j) \cdot S_X \quad (4)$$
$$\Delta y = i \cdot \Delta V_Y \cdot S_Y, \quad (5)$$

accounting for the nonuniform spatial sampling induced by the sinusoidal driving pattern.

To compensate for systematic tilt or alignment errors, global correction terms were added,

$$\Delta x' = \alpha_X \cdot i \quad (6)$$
$$\Delta y' = \alpha_Y \cdot j, \quad (7)$$

where $\alpha_X = 16$ pixels and $\alpha_Y = -16$ pixels are slope coefficients. Fig. 2 illustrate the relative frame relationships under linear and sinusoidal scanning conditions.

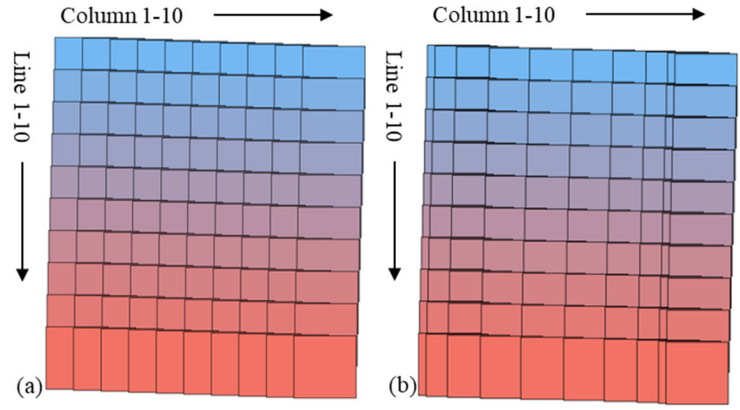

Figure 2. Relative positional relationships between image frames acquired under different scanning strategies. (a) Relative spatial arrangement of image frames under linear scanning. (b) Relative spatial arrangement of image frames under sinusoidal scanning.

The mosaicking process relied on these calibrated scanning parameters to provide input for subsequent image correction and fusion procedures.

## 2.2 Image processing and mosaic reconstruction

The galvanometer-scanned system, introduces spatially dependent intensity variations, such as non-uniform illumination and detector response. During mosaicking, these differences cause perceptible intensity discontinuities at seams degrading visual continuity and hindering subsequent feature extraction. Consequently, additional post-processing is required to ensure brightness consistency.

### 2.2.1 Single-frame intensity correction

Due to the limited effective aperture, each frame contains a circular imaging region and corner brightness-offset regions. To correct this, a fixed rectangular ROI was defined at the offset location. For linear scanning, the ROI was a 580 × 600 pixel region (lower-left). For sinusoidal scanning, two 200 × 400 pixel ROIs (lower-left and lower-right) were selected. A set of reference frames was selected without sample structures were selected, to ensure ROI only reflected the system response.

Assuming a linear response model, the image intensity $I(x, y)$ is expressed as,

$$I(x, y) = g(x, y) \cdot L + o(x, y), \quad (8)$$

where $g(x, y)$ denotes the spatial gain at position $(x, y)$, and $o(x, y)$ represents the additive background offset. The term $L$ corresponds to the global brightness level of the reference frame. Using the bright-field $(I_b, L_b)$ and dark-field $I_g, L_g)$ reference intensities, the response parameters can be solved:

$$g(x,y) = \frac{I_{b(x,y)} - I_{d(x,y)}}{L_b - L_d + \varepsilon} \tag{9}$$

$$o(x,y) = I_d(x,y) - g(x,y) \cdot L_d, \tag{10}$$

where $\varepsilon$ is a small constant introduced to ensure numerical stability. Subsequently, a unified linear correction is applied:

$$I_{corr}(x,y) = \frac{I(x,y) - o(x,y)}{g(x,y) + \varepsilon}. \tag{11}$$

If the brightness-offset region exhibits a significant deviation only under bright-field conditions, the above joint model naturally degenerates into a bright-field-only gain correction:

$$g(x,y) = \frac{I_{b(x,y)}}{L_b + \varepsilon} \tag{12}$$

$$I_{corr}(x,y) = \frac{I(x,y)}{g(x,y) + \varepsilon}. \tag{13}$$

To smooth the transition at the ROI boundaries, a two-dimensional spatial weighting function $W_{ROI}(x,y)$ is constructed within the ROI, decreasing linearly from 1 to 0 within a transition band. The final output image within the ROI is :

$$I_{out}(x,y) = W_{ROI}(x,y) \cdot I_{corr}(x,y) + [1 - W_{ROI}(x,y)] \cdot I(x,y). \tag{14}$$

### 2.2.2 Seam artifact correction

During the mosaicking process, adjacent image frames may exhibit differences in brightness or contrast at the stitching locations. To reduce the perceptual impact of these differences at the seams, a weight-based feathering fusion strategy is applied at the stitching regions between adjacent frames.

Let the overlapping region between two adjacent image frames in the global coordinate system be denoted as $\Omega$. Within this region, a spatial weighting function $W_{seam}(x,y)$ is constructed. The weighting function transitions linearly from 1 to 0 along the stitching direction.

Within the transition region, the fused image intensity is computed as a weighted combination of the two overlapping images:

$$I_{mosaic}(x,y) = W_{seam}(x,y) \cdot I_1(x,y) + [1 - W_{seam}(x,y)] \cdot I_2(x,y), \tag{15}$$

where $I_1(x,y)$ and $I_2(x,y)$ denote the pixel intensities of the two images to be stitched, expressed in the global coordinate system.

Similar to the ROI feathering strategy described in Section 2.2.1, the seam feathering weight varies only within an overlap region of limited width. Its primary role is to smooth low-frequency brightness differences across the stitching boundary, while avoiding significant influence on the high-frequency structural details of the image. Therefore, this approach effectively suppresses seam artifacts while preserving the spatial resolution and fine structural continuity of the mosaicked image.

## 2.3 Quantitative evaluation metrics

To evaluate the mosaicking performance without ground-truth, two complementary aspects were chosen: mosaicking consistency and image quality.

### 2.3.1 Mosaicking consistency metrics

Mosaicking consistency is used to quantify the mean absolute error (MAE) within the overlapping region $\Omega_{ij}$ between the $i$-th image tile $I_i(p)$ and the $j$-th neighboring image tile $I_j(p)$ where $p \in \Omega_{ij}$ [1].

To reduce the influence of global brightness scale differences on the consistency metrics, we applied linear normalization fitting before computing MAE, given as,

$$I_j(p) \approx a \cdot I_i(p) + b \tag{16}$$

$$MAE_{ij} = \frac{1}{|\Omega_{ij}|} \cdot \sum_{p \in \Omega_{ij}} |I_j(p) - (a \cdot I_i(p) + b)|, \tag{17}$$

where the coefficients $a$ and $b$ are estimated within the overlap region using a least-squares fitting procedure.

### 2.3.2 Image quality metrics

The contrast-to-noise ratio (CNR) is used to quantify the feature discriminability, defined as

$$CNR = \frac{|\mu_{sig} - \mu_{bg}|}{\sigma_{bg}}, \qquad (18)$$

where $\mu_{sig}$ and $\mu_{bg}$ denote the mean pixel intensities of the signal and the background region, while $\sigma_{bg}$ represents the standard deviation [8]. For CNR, the signal region is a 400 × 400 pixel black pattern while the background is 1400 × 700 pixel bright area

The image brightness uniformity is evaluated by the intensity standard deviation within the bright region (same as the CNR background) and a 700 x 900 pixel unilluminated dark region. To assess seam artifacts, the average intensity difference between these pixel pairs is computed directly across detected horizontal and vertical seams

## 3. RESULTS

### 3.1 Comparison of raw mosaicking results under linear and sinusoidal scanning

Figures 3(a) and 3(b) show the mosaicking results obtained under linear scanning and sinusoidal scanning without applying any additional image processing methods. For both scanning strategies, a 30 ms settling time was applied at each position, with a total of 10 × 10 frames were acquired to cover 2.5 × 2.5 cm² within a total acquisition time of 6.05 s. Based on the resolution target features, the maximum achievable resolution under both mosaicking strategies is measured to be 7.81 µm, and this resolution is maintained uniformly across the entire FOV. The quantitative mosaicking consistency metrics introduced in Section 2.3 are summarized in Table 1. Notably, sinusoidal scanning exhibits a larger MAE within the overlap regions compared with linear scanning, indicating a higher degree of pixel-level mismatches in the raw data. This result suggests that the unprocessed linear scanning provides better controllability and higher consistency in geometric mosaicking.

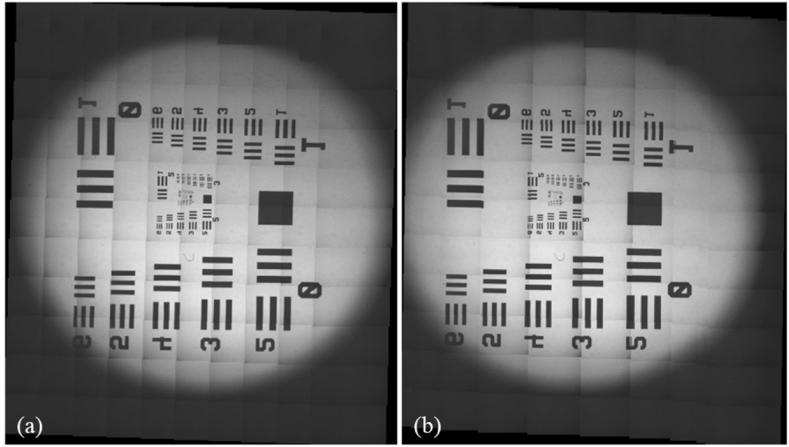

(a) (b)

Figure 3. Raw mosaicking results obtained under different scanning strategies without image processing. (a) Linear scanning. (b) Sinusoidal scanning.

Table 1. Quantitative comparison of mosaicking consistency and image quality metrics for linear and sinusoidal scanning, before and after image processing.

| Metric | Linear (Raw) | Linear (Processed) | Sinusoidal (Raw) | Sinusoidal (Processed) |
| --- | --- | --- | --- | --- |
| MAE (overlap region) | 248.1622 | | 400.8950 | |
| Bright-region std ($\sigma_{bg}$) | 471.7634 | 417.9232 | 478.0795 | 476.8066 |
| Dark-region std ($\sigma_{bg}'$) | 79.7187 | 77.4591 | 84.7707 | 83.8093 |
| CNR | 7.0702 | 7.1416 | 7.3108 | 7.3302 |
| Mean seam jump | 303.8861 | 235.1487 | 271.1985 | 263.2314 |

## 3.2 Image quality evaluation of linear and sinusoidal mosaicking after processing

As summarized in Table 1, the CNR increases for both scanning strategies after processing. Specifically, the CNR of linear mosaicking increases from 7.07 to 7.14, while the CNR of sinusoidal mosaicking increases from 7.31 to 7.33. Although the absolute improvement in CNR is moderate, the consistent increase across both scanning modes indicates an overall enhancement in contrast preservation after brightness correction. In terms of brightness uniformity, the standard deviation of pixel intensity in both bright and dark regions decreases after processing for linear scanning, reflecting improved intensity consistency within homogeneous regions. For sinusoidal scanning, the corresponding reductions are less pronounced than those observed under linear scanning, indicating that under nonuniform spatial sampling, the effectiveness of brightness adjustment based on the same ROI is partially limited.

A more substantial improvement is observed in the seam-related metrics. For linear mosaicking, the mean seam brightness jump is reduced markedly from 303.9 to 235.1, indicating a significant suppression of brightness discontinuities at stitching boundaries. In contrast, sinusoidal mosaicking exhibits a more modest reduction in seam brightness jump, decreasing from 271.2 to 263.2. These results suggest that the adopted seam balancing and feathering strategies are particularly effective under linear scanning, where the spatial sampling and overlap geometry are more uniform. Notably, despite the reduction of seam artifacts, fine structural details within the mosaicked images do not exhibit noticeable degradation. Although faint seam-related linear features may still be discernible, the absence of obvious blurring or loss of high-frequency content indicates that the proposed processing approach effectively mitigates seam artifacts while preserving high-frequency image information.

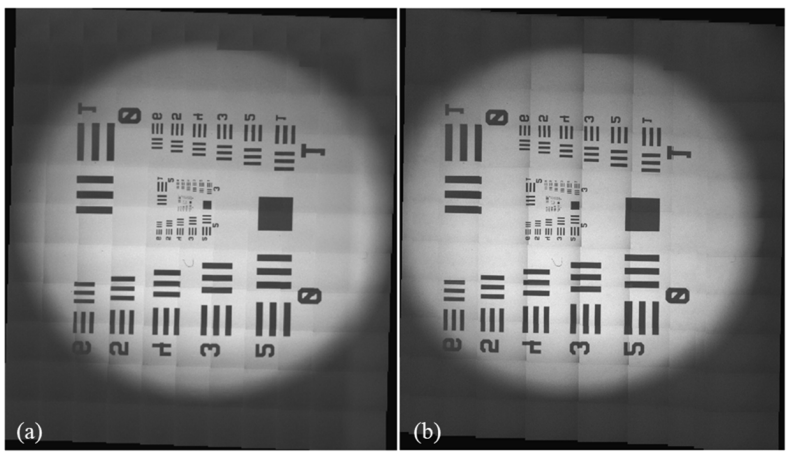

Figure 4. Processed mosaicking results under linear and sinusoidal scanning after brightness correction and seam balancing. (a) Linear scanning. (b) Sinusoidal scanning.

## 4. CONCLUSIONS

In this study, we presented an image mosaicking framework for wide-field microscopic imaging that is applicable to both linear and sinusoidal galvanometric scanning strategies. A translation-based mosaicking framework was combined with ROI-based brightness correction and seam artifact correction to improve image consistency across large fields of view.

Using the proposed geometric mosaicking strategy, wide-field images were successfully reconstructed for both linear and sinusoidal scanning. Quantitative evaluation showed that sinusoidal scanning exhibited larger pixel-level mismatches in overlap regions compared with linear scanning due to nonuniform spatial sampling. After image processing, both scanning strategies showed improved image quality, including enhanced brightness uniformity, increased contrast-to-noise ratio, and reduced seam-related artifacts. The reduction in seam brightness discontinuities was more pronounced for linear mosaicking, while fine structural details and high-frequency image information were largely preserved.

These results highlight the importance of jointly considering scanning strategy and post-processing design for wide-field microscopic mosaicking. The presented analysis and processing framework offer valuable insights into selecting scanning schemes and designing image correction pipelines in scan-based imaging systems.